\documentclass[aps,prl,twocolumn,groupedaddress]{revtex4}

\usepackage{graphicx}
\usepackage{color}
\usepackage{dcolumn}
\usepackage{amsmath}
\usepackage{amssymb}

\newcommand{\kag}{kagom\'{e}}

\begin{document}

\title{Ultracold Atoms in a Tunable Optical Kagom\'{e} Lattice}

\author{Gyu-Boong Jo$^1$}
\author{Jennie Guzman$^1$}
\author{Claire K. Thomas$^1$ }
\author{Pavan Hosur$^1$}
\author{Ashvin Vishwanath$^{1,2}$}
\author{Dan M. Stamper-Kurn$^{1,2}$}


\affiliation{$^1$Department of Physics, University of California, Berkeley CA 94720    \\
$^2$Materials Sciences Division, Lawrence Berkeley National Laboratory, Berkeley, CA 94720}
\date{\today}


\begin{abstract}
Geometrically frustrated systems with a large degeneracy of low energy states are of central interest in condensed-matter physics \cite{rami94review,bale10review}.  The \kag\ net -- a pattern of corner-sharing triangular plaquettes -- presents a particularly high degree of frustration, reflected in the non-dispersive orbital bands.  The ground state of the \kag\ quantum antiferromagnet, proposed to be a quantum spin liquid or valence bond solid \cite{else89,mars91,sach92,nico03kagome,wang06projective,ran07projected,sing07ground,yan11kagome}, remains uncertain despite decades of work.  Solid-state \kag\ magnets \cite{shor05herb,hiro01kagome}
suffer from significant magnetic disorder or anisotropy that complicates the interpretation of experimental results.  Here, we realize the \kag\ geometry in a two-dimensional optical superlattice for ultracold $^{87}$Rb atoms.  We employ atom optics to characterize the lattice as it is tuned between various geometries, including \kag, one-dimensional stripe, and decorated triangular lattices, allowing for a sensitive control of frustration.  The lattices implemented in this work offer a near-ideal realization of a paradigmatic model of many-body quantum physics.
\end{abstract}

\maketitle


Ultracold atoms trapped within optical lattices exhibit correlated phases, such as Mott insulators of bosons \cite{grei02mott} and fermions \cite{jord08mott} and effective antiferromagnetic ordering in one-dimensional chains \cite{simo11qsim}, which are analogous to those studied in condensed-matter physics.  While most experiments have been performed in primitive Bravais lattices, with a single lattice site per unit cell, recent works have explored non-standard optical lattices with a few-site basis such as the two-dimensional honeycomb \cite{solt11hexagonal} and checkerboard \cite{wirt11pband} geometries, and double-well superlattices \cite{sebb06,foll07second}.  Such lattices add a low-energy orbital degree of freedom, corresponding to the assignment of a particle to each site within the unit cell, that can yield non-trivial ordering and dynamics.

The \kag\ lattice is obtained by eliminating every fourth site from a triangular lattice of spacing $a/2$, with the eliminated sites forming a triangular lattice of spacing $a$.  The remaining sites generate three connected $s$-orbital bands within a bandwidth on the order of the intersite tunneling energy. Intriguingly, the frustration of antiferromagnetic interactions in the \kag\ geometry implies that one of these  bands be non-dispersing.  Such flat bands accentuate the role of interparticle interactions, leading possibly to crystalline ordering \cite{wu07flat} and supersolidity \cite{hube10flat} for scalar bosons, and ferromagnetism of itinerant fermions \cite{tasa92}.

\begin{figure*}[t]
\centering
\includegraphics[width=6.8in]{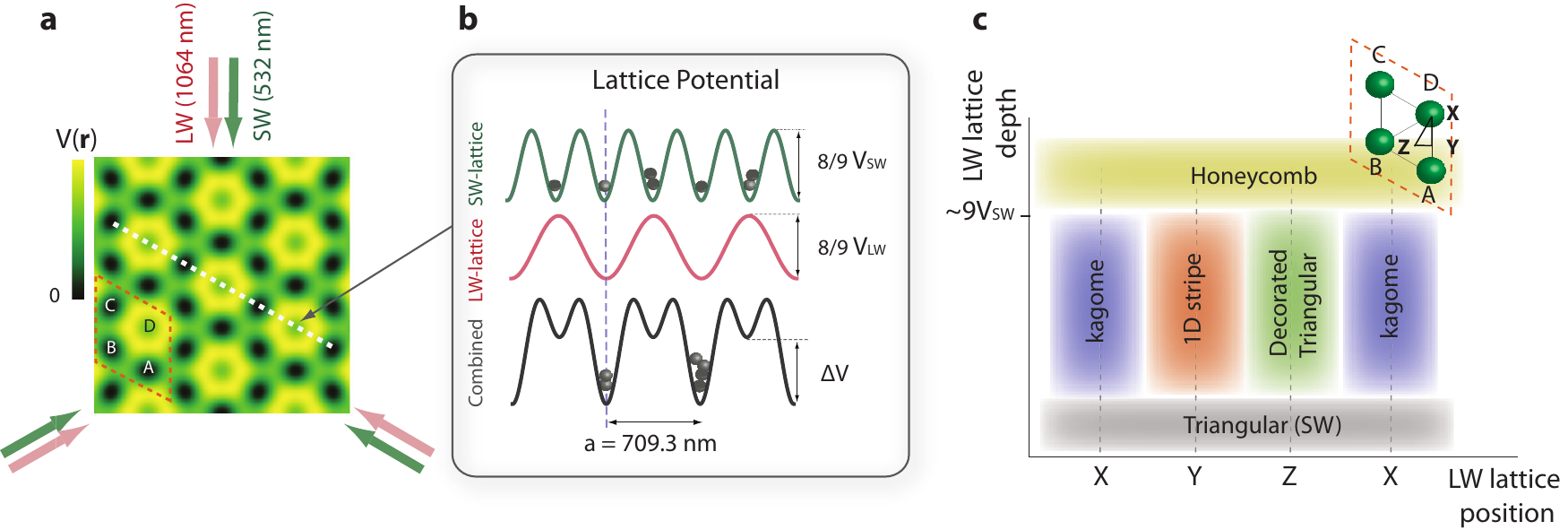}
\caption{\textbf{Tunable optical \kag\ lattice.} Three bichromatic light beams intersecting at 120$^\circ$ form a \kag\ optical lattice for ultracold $^{87}$Rb atoms, with the two-dimensional optical potential $V(\mathbf{r})$ shown in (a).  Profiles of the potential of the SW, LW, and combined lattices are shown in (b).  Sites D of the SW lattice are emptied as their energy offset $\Delta V$ exceeds the gas chemical potential, so that the remaining sites A, B and C form the \kag\ geometry.  (c) Different lattice geometries are created for intermediate LW lattice depths ($V_{LW} < 9 V_{SW}$)) by displacing the potential maxima of the SW lattice to the high-symmetry points X, Y or Z within the unit cell, while, for higher LW lattice depths, a honeycomb geometry prevails.}
\label{fig:scheme}
\end{figure*}

Here, we realize a two-dimensional \kag\ lattice for ultracold atoms by overlaying two commensurate triangular optical lattices generated by light at the wavelengths of 532 and 1064 nm (Fig.\ \ref{fig:scheme}).  Stabilizing and tuning the relative position of the two lattices, we explore different lattice geometries including a \kag\ lattice, a one-dimensional (1D) stripe lattice, and a decorated triangular lattice.  We characterize these geometries using Kapitza-Dirac (KD) diffraction and by analyzing the Bloch-state composition of a superfluid released suddenly from the lattice.  The Bloch-state analysis also allows us to determine the ground-state distribution within the superlattice unit cell as the potential is tuned between geometries.  The atom-optical tools developed in this work are useful for the characterization of other superlattices and for superlattice-based atom interferometry.

Three plane waves of light of equal intensity $I$, equal frequency, and wavevectors (and linear polarizations) lying in a plane and intersecting at equal angles, produce a triangular lattice of points with zero intensity, and a honeycomb lattice of points with maximum intensity $\frac{9}{2} I$ separated by a triangular lattice of intensity saddle points with intensity $4 I$.  In our work, the short-wavelength (SW) lattice is formed by light that is blue-detuned from the principal atomic resonances of rubidium, so that atoms are attracted to the triangular lattice of zero-intensity sites with a lattice spacing of $a/2 = (2/3)\times 532 \, \mbox{nm} = 355 \, \mbox{nm}$.  The long-wavelength (LW) lattice is red detuned, so that its zero-intensity points are potential-energy maxima for rubidium atoms.  A unit cell of the LW lattice contains four sites of the SW triangular lattice, labeled A,B,C and D in Fig.\ \ref{fig:scheme}.  Aligning the positions of the LW potential maxima to coincide with sites D lowers the potential energies $V_{A,B,C}$ at the indicated sites by equal amounts $\Delta V = V_D - V_{A,B,C} = \frac{8}{9} \, V_{LW}$ where $V_{LW}$ is the maximum scalar potential depth of the LW lattice (we ignore the $\sim 1 \%$ vector shift in this lattice \cite{solt11hexagonal}).  As $\Delta V$ is increased, atoms are excluded from sites D, while the remaining sites form the \kag\ optical lattice.  The \kag\ geometry persists until $V_{LW} > 9 \, V_{SW}$, at which point atoms become preferentially confined in the LW honeycomb lattice.

Compared with previous proposals \cite{sant04kagome,ruos09kag}, our simpler approach to creating a \kag\ lattice allows one to tune the lattice geometry, thereby controlling its degree of frustration.  Aligning the LW potential maxima with the SW lattice saddle points disfavors population in two sites of the four-site unit cell (e.g.\ $V_{B,C} < V_{A,D}$) producing a 1D stripe lattice (Fig.\ \ref{fig:scheme}c).  Aligning the LW potential maxima with the SW potential maxima disfavors population in three sites of the four-site unit cell (e.g.\ $V_{A,B,D}>V_C$), producing a decorated triangular lattice with lowest-energy sites forming a triangular lattice while the remaining sites form a \kag\ lattice of local potential minima.

Experiments were conducted with scalar Bose-Einstein condensates (BECs) of $\sim3\times10^{5}$ $^{87}$Rb atoms produced at temperatures of about 80 nK in a red-detuned crossed optical dipole trap with trap frequencies of $(\omega_x,\omega_y,\omega_z)=2\pi\times(60,30,350) \, \mbox{Hz}$, with $\omega_z$ applying vertically.  The relative position of the LW and SW lattices was measured using two separate two-color interferometers and stabilized actively by displacing mirrors in the relevant optical paths.  The relative position was tuned along two linearly independent directions by adding within each interferometer a variable relative phase shift between the two lattice colors (see Methods).


The first atom-optical method we used to demonstrate the tunable superlattice is Kapitza-Dirac diffraction \cite{goul86,ovch99}, for which the lattice potential is suddenly switched on and off, after which the condensate is released from the optical trap and imaged after a time of flight to evaluate its momentum-space distribution.  The pulse is sufficiently short that one can neglect the kinetic energy of the diffracted momentum components over the pulse duration $\tau$.  The effect of the pulse is then to imprint a phase $-V(\mathbf{r}) \tau/ \hbar$ proportional to the lattice potential $V(\mathbf{r})$ onto the condensate wavefunction (which is initially nearly uniform).

\begin{figure}
\begin{center}
\includegraphics{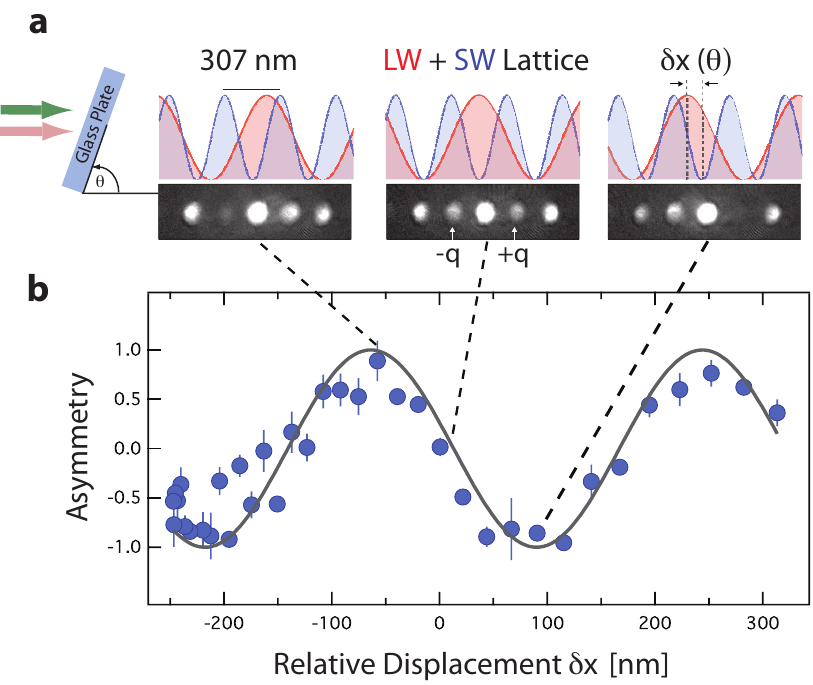}
\caption{\textbf{Kapitza-Dirac diffraction from a 1D superlattice.}  Atom diffraction patterns, formed by $\tau=\mbox{8}$~$\mu$s pulse of the lattice potential followed by 26~ms time of flight, exhibit left/right momentum asymmetry (defined as $(P_{+q}-P_{-q})/(P_{+q}+P_{-q})$ that varies as the displacement $\delta x$ between the LW- and SW-lattice intensity minima is varied, reflecting broken inversion symmetry of the lattice, in close agreement with the predicted behavior (solid line).} 
\label{fig:kd}
\end{center}
\end{figure}

The corresponding momentum-space distribution is sensitive to the relative displacement of the LW and SW lattices. To exhibit this sensitivity we blocked one of the incident bi-chromatic lattice beams and examined the resulting one-dimensional superlattice formed by two bichromatic light beams intersecting at an angle of  120$^{\circ}$, with potential energy given as  $V(x) = V_{LW} \sin^2( q (x + \delta x)/2 ) - V_{SW} \sin^2(q x)$ where $2 \pi / q = 614 \, \mbox{nm}$ is the 1D LW lattice spacing, and $\delta x$ is the distance between the LW and SW intensity minima.  Considering diffraction up to second order in the phase modulation depth, the atomic populations at wavevectors $\pm q$ are given as\begin{equation}
P_{\pm q} \propto \left| \pm i J^{LW}_{\pm 1} J^{SW}_0 + J^{LW}_{\mp 1} J^{SW}_{\pm 1} e^{\mp i 2 q \delta x} \right|^2
\end{equation}
where $J_n$ is the $n^{\mbox{th}}$-order Bessel function, evaluated at $\phi_{LW,SW} = V_{LW,SW} \, \tau/2 \hbar$ according to the superscript.  The lack of inversion symmetry of the lattice produced by an incommensurate value of $\delta x$ appears as a left/right momentum asymmetry in the diffracted matter wave (Fig.\ \ref{fig:kd}). 


A second method to characterize the optical superlattice is the momentum-space analysis of a superfluid occupying the ground state of the lattice potential.  For such analysis, the optical lattice potential depth was ramped up from zero over 90~ms, held at a constant level for 100~ms, and then suddenly switched off. Imaging after a time of flight reveals coherent momentum peaks of the condensate wavefunction at the reciprocal lattice vectors $\mathbf{G}$ of the LW lattice.

\begin{figure}
\includegraphics{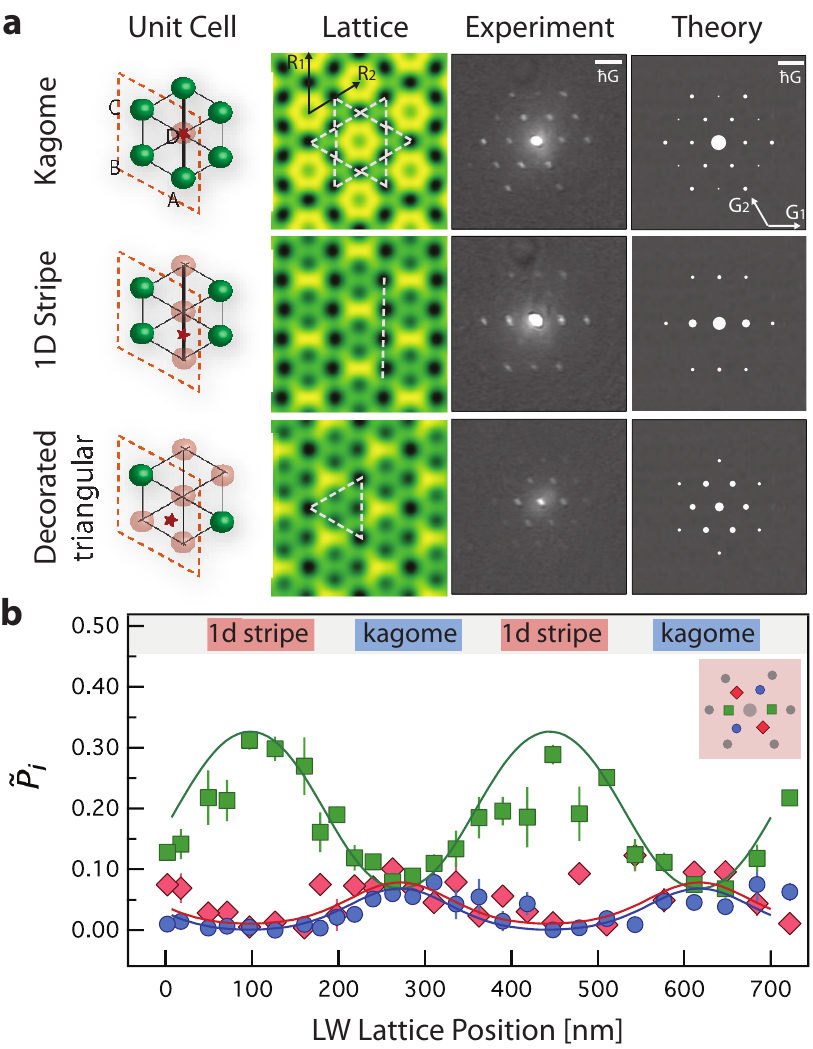}
\caption{\textbf{The momentum-space composition of a BEC for various lattices.} (a) The \kag\ and decorated triangular lattices  maintain three-fold rotational symmetry in configuration and momentum space, while the symmetry of the 1D stripe lattice is reduced to a parity symmetry (left-right in the images).  For each setting, a schematic distinguishes between sites of high (green) and low (red) atomic population.  The lattice potential and expected momentum distribution are calculated from measured values of the LW and SW lattice depths. The area of the white dot reflects the momentum population in the expected diffraction pattern. (b) Translating the LW-lattice potential maxima along the axis connecting sites A and D tunes the lattice between \kag\ and 1D stripe geometries, as revealed by the population ratios $\tilde{P}_i$ identified by color according to the inset.  The data (averages over 4-5 repeated measurements) agree with calculations of the single-particle ground state (solid lines) with $\Delta V=14\, \mbox{kHz}$ and $V_{SW}=40\, \mbox{kHz}$ measured by Kapitza-Dirac scattering from LW-only or SW-only lattices.  $\Delta V$ was higher than the chemical potential $\mu \sim h \times 3.5 \, \mbox{kHz}$ of the condensate in the SW-only lattice, so that modifications of the transverse condensate wavefunction due to interactions are negligible.} \label{fig:bragg}
\end{figure}

\begin{figure}
\includegraphics{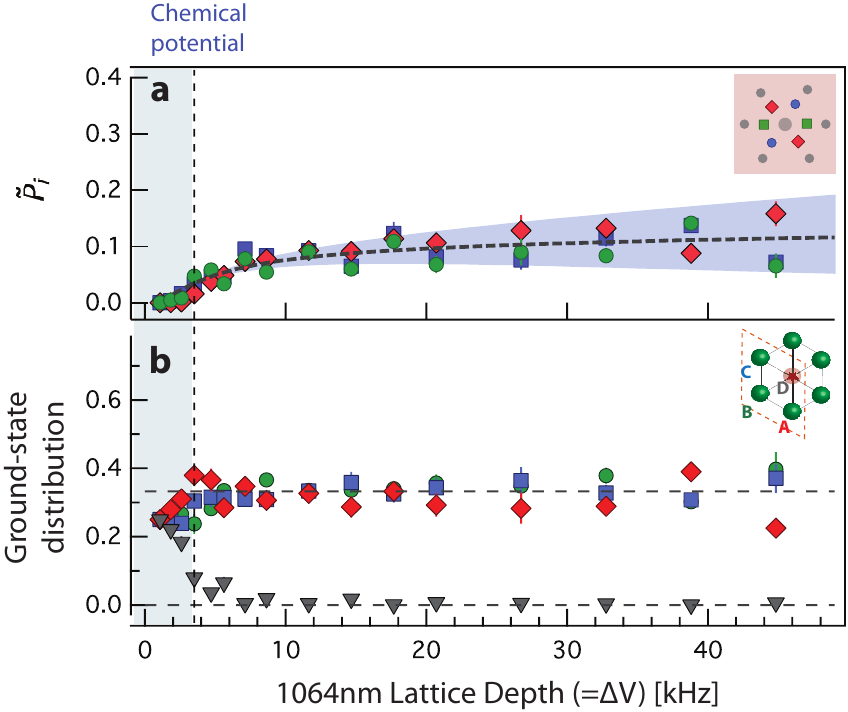}
\caption{ \textbf{Momentum population ratios and atomic distributions within the unit cell for a \kag\ lattice.} The superlattice was converted from a SW triangular to a \kag\ lattice by increasing $V_{LW}$. The D-site population is extinguished (b) as the energy offset $\Delta V$ exceeds the condensate chemical potential ($\mu \simeq h \times 3.5 \, \mbox{kHz}$), and the momentum population ratios reach the asymptotic value of 1/9 expected for a \kag\ lattice (a).  Data points represent averages of 7 -- 10 measurements.  Dashed curves indicate expected population ratios $\tilde{P}_i$ for the ideal \kag\ configuration while the shaded region indicates the expected variation in $\tilde{P}_i$ given a shot-to-shot instability of $\sim$20~nm in the relative position of the LW and SW lattices.}\label{fig:unitcellpop}
\end{figure}

Varying the relative position of the two lattices we identify the three high-symmetry lattice configurations (Fig.\ \ref{fig:bragg}a).  Given that the scalar condensate occupies the ground state of the lattice potential, its wavefunction can be taken as real and positive; thus, its momentum distribution is symmetric under inversion.  Expansion from both the \kag\ and the decorated triangular lattices shows the three-fold rotational symmetry of the optical superlattice.  In the 1D stripe geometry, one expects equally weak occupation of two sites (e.g., A and D), and equally strong occupation of the other two sites (B and C) of the superlattice unit cell.  Such a distribution is (nearly) invariant under displacements of $a/2$ along the A-D axis, and condensate diffraction along that axis should reflect the shorter periodicity of the SW lattice.  The momentum distribution should also be symmetric under reflection about the A-D axis. Both traits are observed experimentally.

The Bloch-state momentum distributions allow one to quantify the ground-state wavefunction within a unit cell of the superlattice, which we express as $\psi(\boldsymbol{r})=\sum_{\alpha} \psi_\alpha w_{\alpha}(\boldsymbol{r}-\boldsymbol{s}_{\alpha})$ where $w_{\alpha}(\boldsymbol{r})$ is the normalized Wannier state wavefunction, $\boldsymbol{s}_{\alpha}$ the position and $|\psi_\alpha|^2$ the fractional atomic population of site $\alpha\in\{A,B,C,D\}$ of the unit cell.  At low $V_{LW}/V_{SW}$, we approximate $w_{\alpha}=w$ as cylindrically symmetric and identical for all $\alpha$.
From the momentum-space populations $P_{\mathbf{G}_i}$ ($i \in \{1, 2, 3\}$) in the three first-order diffraction peaks of the LW lattice -- corresponding to the inner hexagon of peaks in time-of-flight images -- and that at zero wavevector $P_{\mathbf{0}}$, one determines the distinct quantities
\begin{equation}
\tilde{P}_{i}=\frac{P_{\mathbf{G}_{i}}+P_{-\mathbf{G}_{i}}}{2 P_{\mathbf{0}}} \frac{|w(0)|^{2}}{|w(\mathbf{G}_i)|^{2}}=\frac{\left| \psi_\beta + \psi_\gamma - \psi_\delta - \psi_\epsilon\right|^{2}}{\left|\sum_{\alpha} \psi_\alpha \right|^{2}}\label{eq:firstorder}
\end{equation}
where $w(\mathbf{0})$ and $w(\mathbf{G}_i)$ are now Fourier components of the Wannier function, and $\beta$, $\gamma$, $\delta$ and $\epsilon$ label the four sites so that $\mathbf{G}_{i}\cdot(\mathbf{s}_{\beta}-\mathbf{s}_{\gamma})=0$. Together with the normalization $\sum_{\alpha} |\psi_{\alpha}|^2=1$  these quantities determine the atomic distribution in the unit cell.   We note that inverting Eqs.\ \ref{eq:firstorder} gives a discrete degeneracy of solutions that can be resolved by prior knowledge of the approximate populations.

We measured the population ratios $\tilde{P}_{i}$ as the superlattice geometry was gradually tuned, using data from the second-order diffraction peaks to characterize the Wannier state (see Methods).  Translating the relative position of the two lattices (Fig.\ \ref{fig:bragg}b), one advances from the \kag\ geometry, with equal population in the three ratios, to the 1D stripe geometry, with two identically small ratios, and then to another \kag-geometry lattice.  Our data agree with a calculation of the ground-state single-particle wavefunction for the known lattice depths.

We focus finally on the \kag-geometry lattice alignment, and examine the transition between the triangular and \kag\ geometries (Fig.\ \ref{fig:unitcellpop}).  At zero LW lattice depth, the atoms are confined in a SW triangular lattice, and the first-order LW lattice diffraction orders are absent, indicating a unit-cell population of $(A,B,C,D)=(1/4,1/4,1/4,1/4)$.  As the LW lattice depth is increased,  the population ratios $\tilde{P}_i$ increase and the \kag\ geometry is achieved by gradually expelling atoms from one site of the unit cell.  The population ratios tend toward a limiting value of 1/9 that is a hallmark of diffraction from a \kag\ lattice wherein the atoms are distributed as $(A,B,C,D)=(1/3,1/3,1/3,0)$ (see Methods).

Here, the ground state of the \kag\ lattice does not suffer from frustration, allowing for the stability of a Bose-Einstein condensate in the lattice.  In the future, effects of frustration may be explored by transferring
bosons into the excited $s$-orbital flat band, or by changing the sign of the hopping energy \cite{lign07shake} so that the flat band becomes the lowest-energy band.  We note that the present choice of wavelengths also yields \kag\ lattices for the fermionic isotopes  $^6$Li and $^{40}$K.  Introducing fermions into the lattice at the appropriate fillings would place the Fermi energy within the flat band, allowing for studies of flat-band ferromagnetism due to repulsive interactions \cite{tasa92} or of enhanced Cooper pairing for attractive interactions \cite{imad00}. Furthermore, remarkable versatility of the 2D superlattice
opens new possibilities to emulate a distorted \kag\ lattice as well as a nearly ideal one enabling us to explore
the possible various quantum ground states of the \kag\ quantum antiferromagnet.

\section{ Methods}

\paragraph{\textbf{Optical setup}}
Lattice beams with Gaussian spatial profiles, lying and polarized in the horizontal plane, were centered onto the condensate.  The large $\sim 100 \, \mu\mbox{m}$ beam-waist diameters of the lattice beams ensured that the trapping frequencies were not strongly modified by the lattice potential.  Laser alignment and relative intensities were tuned to produce six-fold symmetric diffraction patterns of condensates released from LW- and SW-only lattices.  In the combined lattice, the relative displacement of the LW and SW lattices was measured using two two-color Mach-Zehnder interferometers, one for beams 1 and 2 and the other for beams 1 and 3, and stabilizing all four interference signals by feedback to separate piezo-actuated mirrors in the optical paths.  A rotatable glass plate within each interferometer introduces a controlled relative shift between the two lattice colors that, following stabilization, is then imparted onto the lattice beams at the location of the optical lattice.  For the Kaptiza-Dirac diffraction experiments, the active stabilization was engaged with the lattice light at negligible intensity before the brief high-intensity pulse of the lattice potential.  For the Bloch-state analysis experiments, the stabilization was involved at the beginning of the ramp-on of the lattice beams.  To reduce deleterious effects of atomic interactions in the time-of-flight analysis, the vertical optical confinement was switched off about 0.5 ms before all other optical trapping beams, including the lattice beams, were extinguished.

The optical frequencies of the SW and LW lattice beams were stable within a few tens of MHz, at the wavelengths of 532.15~nm and 1063.96~nm respectively.  While the wavelengths are not perfectly commensurate, we note that the relative displacement of the SW and LW lattices varies by only (18,9)~nm over the $(d_x,d_y)=(28,14)\,  \mu\mbox{m}$ transverse diameter of the condensates in our experiments.

\paragraph{\textbf{Bloch-state analysis}}
Momentum populations at the reciprocal lattice wavevectors are measured by integrating under the narrow atomic distributions observed in time of flight images, thereby accounting for effects of finite temperature, atomic interactions, and finite size during the expansion of the gas.  The Wannier state Fourier components that appear in Eq.\ \ref{eq:firstorder} are determined, assuming the Wannier state to be Gaussian, from the second-order diffraction populations as $|w(\mathbf{0})|^2 / |w(\mathbf{G}_i)|^2 = \sqrt{2 P_{\mathbf{0}} / (P_{2 \mathbf{G}_i} + P_{- 2 \mathbf{G}_i})}$. The limiting values of the momentum population ratios enable us to distinguish different lattice geometries as summarized in Table  \ref{tab:bragg ratio main} .

\begin{table}
\begin{centering}
\begin{tabular}{|c|c|}
\hline
Lattice geometry& Momentum population ratio $ \tilde{P}_i$ \tabularnewline
\hline
\hline
Triangular (SW) &$0$ \tabularnewline
\hline
\kag\ & $1/9$\tabularnewline
\hline
1D stripe & $1$\tabularnewline
\hline
Decorated triangular & $1$\tabularnewline
\hline
Honeycomb & $1/4$\tabularnewline
\hline
\end{tabular}
\par\end{centering}

\caption{The momentum population ratios for various lattice geometries assuming the Wannier wavefunctions at all sites to be $\delta$-functions in real space. The hallmark of the \kag\ lattice is the limiting value of 1/9 as observed in Fig.\ \ref{fig:unitcellpop}.  Although the 1D stripe and the decorated
triangular lattices have the same momentum ratios in this table,
they can be easily distinguished through their rotational symmetries,
as can be seen in Fig. \ref{fig:bragg}.\label{tab:bragg ratio main}}

\end{table}

\section{acknowledgement}

We thank  S.\ Schreppler for experimental assistance.  C.\ K.\ Thomas acknowledges support by the Department of Energy Office of Science Graduate Fellowship Program (DOE SCGF), made possible in part by the American Recovery and Reinvestment Act of 2009, administered by ORISE-ORAU under contract DE-AC05-06OR2310. P. H. was supported by NSF-DMR 0645691. This work was supported by the NSF and by the Army Research Office with funding from the DARPA Optical Lattice Emulator program.


\end{document}